\documentclass{PoS}
\def\snn{${ \sqrt{s_{_{NN}}}~ = \rm {200~GeV} }$}

\def\pt{$p_{\rm T}$ }

\def\pp{$p$+$p$ }

\title{Silicon Vertex Tracker for PHENIX Upgrade at RHIC: Capabilities
and Detector Technology} \ShortTitle{Vertex Tracker for PHENIX Upgrade
at RHIC} \author{Rachid Nouicer for the PHENIX Collaboration\\ Physics
Department, Brookhaven National Laboratory,\\ Upton, New York
11973-5000, U.S.A.\\ E-mail: \email{rachid.nouicer@bnl.gov}}
\abstract{From the wealth of data obtained from the first three years
of RHIC operation, the four RHIC experiments, BRAHMS, PHENIX, PHOBOS
and STAR, have concluded that a high density partonic matter is formed
at central Au+Au collisions at \snn. The research focus now shifts
from initial discovery to a detailed exploration of partonic matter.
Particles carrying heavy flavor, i.e. charm or beauty quarks, are
powerful tool for study the properties of the hot and dense medium
created in high-energy nuclear collisions at RHIC. At the relatively
low transverse momentum region, the collective motion of the heavy
flavor will be a sensitive signal for the thermalization of light
flavors. They also allow to probe the spin structure of the proton in
a new and precise way. An upgrade of RHIC (RHIC-II) is intended for
the second half of the decade, with a luminosity increase to about
20-40 times the design value of 8 $\times$ 10$^{\rm 26}$ cm$^{\rm -2}$
s$^{\rm -1}$ for Au+Au, and 2 $\times$ 10$^{\rm 32}$ cm$^{\rm -2}$
s$^{\rm -1}$ for polarized proton beams. The PHENIX collaboration
plans to upgrade its experiment to exploit with an enhanced detector
new physics then in reach. For this purpose, we are constructing the
Silicon Vertex Tracker (VTX). The VTX detector will provide us the
tool to measure new physics observables that are not accessible at
the present RHIC or available only with very limited accuracy.  These include a
precise determination of the charm production cross section,
transverse momentum spectra at high-${\rm p_{_{T}}}$ region for
particles carrying beauty quarks as well the detection of recoil jets
in direct photon production. The VTX detector consists of four layers of
barrel detectors located in the region of pseudorapidity ${\rm
|\eta|<}$~1.2 and covers almost 2$\pi$ azimuthal angle. 
The pseudorapidity, ${\rm \eta }$, is defined as
${\rm \eta = -ln[tan(\theta/2)]}$, where ${\rm \theta }$
is the emission angle relative to the beam axis. The inner two
silicon barrels consists of silicon pixel sensors and their technology
is the ALICE1LHCb sensor-readout hybrid, which was developed at CERN
for the ALICE and LHCb experiments.  The outer two barrels consists of
silicon stripixel detector with a new "spiral" design, single-sided
sensor with 2-dimensional (X-U) read-out.  In this paper, we will
provide details of the physics capability added to PHENIX by the new
central silicon vertex tracker, the status of the project, including
technology choices used in the design, performance of individual
silicon sensor and silicon detector prototype.}  \FullConference{The
16$^{\rm th}$ International Workshop on Vertex detectors\\ September
23-28 2007\\ Lake Placid, NY, USA}
\begin{document}
\begin{figure}[ht]
\centering
\vspace*{-0.3cm}
\includegraphics[height=.3\textheight]{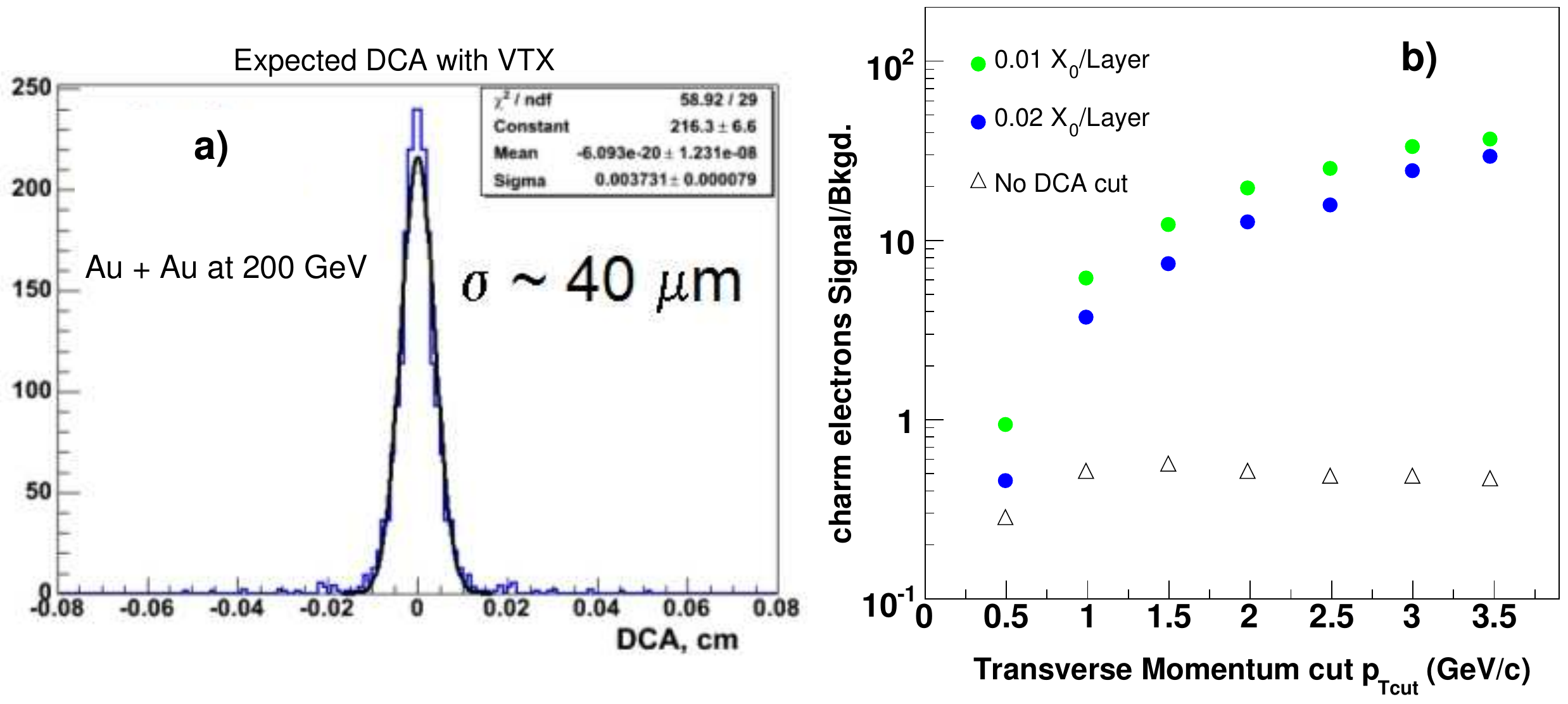}
\vspace*{-0.4cm}
\caption{\small Panel a): DCA distribution for 3 $\le$ $\rm p_{T}$
$\le$ 4 GeV/c pions in the PISA simulation of the VTX detector. The
DCA resolution of $\sigma \sim $ 40 $\mu$m was achieved using the two
inner pixel barrels. Panel b): Signal to Background ratios as a
function of minimum electron \pt\ cut. The signal corresponds to
electrons issued from charm decays and they are obtained using a DCA
cut of 200 $\mu$m (dots) or no DCA cut (triangles). The
background corresponds to electrons from Dalitz decays and photon
conversions which pass the corresponding DCA cuts, assuming four
layers of silicon with 1 or 2\% of a radiation length per layer.
\label{fig:fig1}}
\end{figure}
\begin{figure}[ht!]
\centering
  \begin{tabular}{cc}
\\
    \begin{minipage}{3in}
  \hspace*{-2.0cm}
\vskip -1cm \includegraphics[height=.42\textheight]{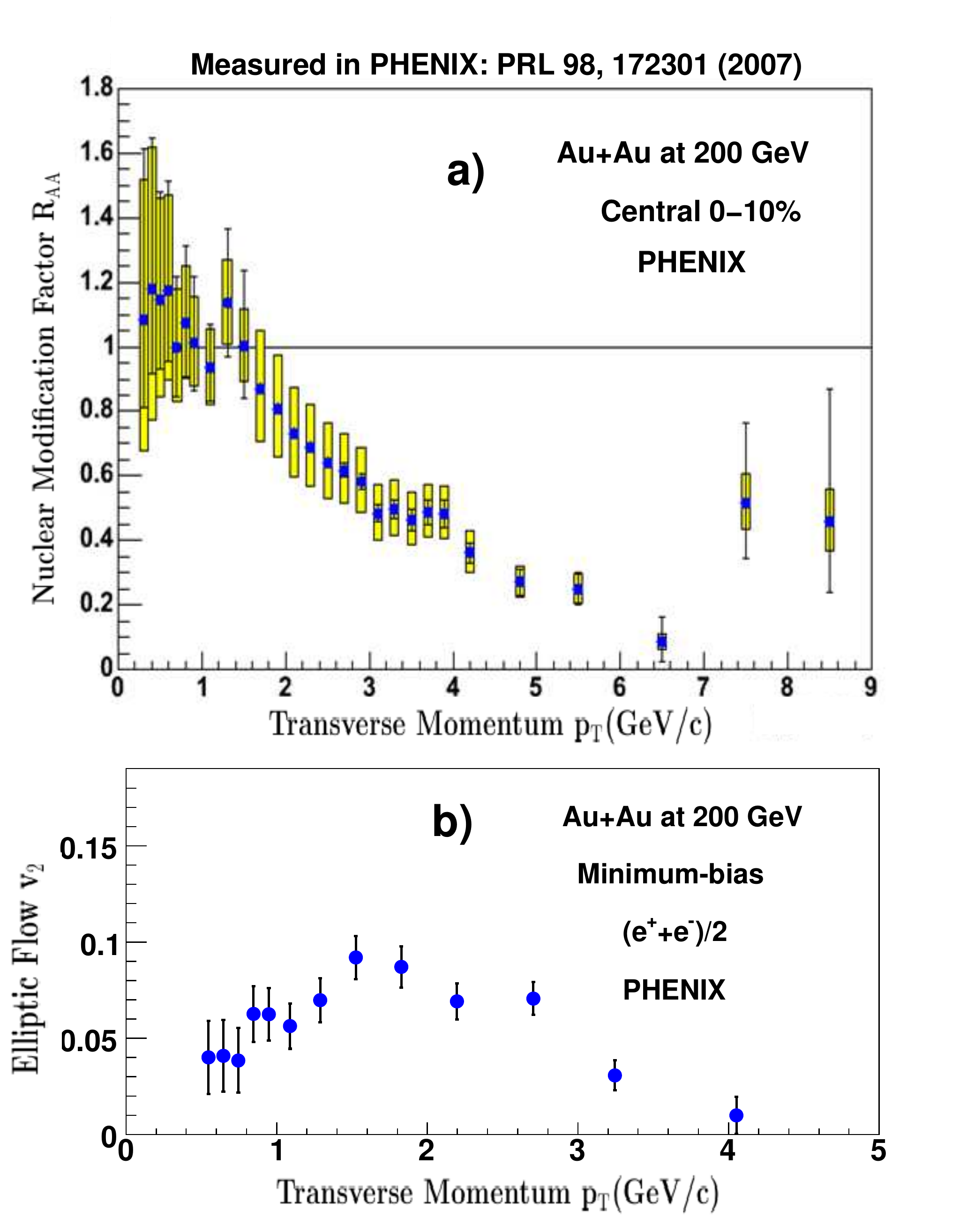}
\vspace*{-0.2cm}
\caption{\small Measured: a) nuclear modification factor, $R_{\rm AA}$,
and b) elliptic flow, ${\rm v_2}$, as a function of transverse
momentum of heavy-flavor electrons, $(e^{+} + e^{-}$)/2, from
heavy-flavor decays at midrapidity region in Au+Au collisions at
\snn. These measurements have been obtained by present PHENIX
experiment~\cite{PHENIX1}.
\label{fig:fig2}}
    \end{minipage}
&\hspace*{-0.1cm}
    \begin{minipage}{3.in}
\vspace*{-0.7cm}
\includegraphics[height=.4\textheight]{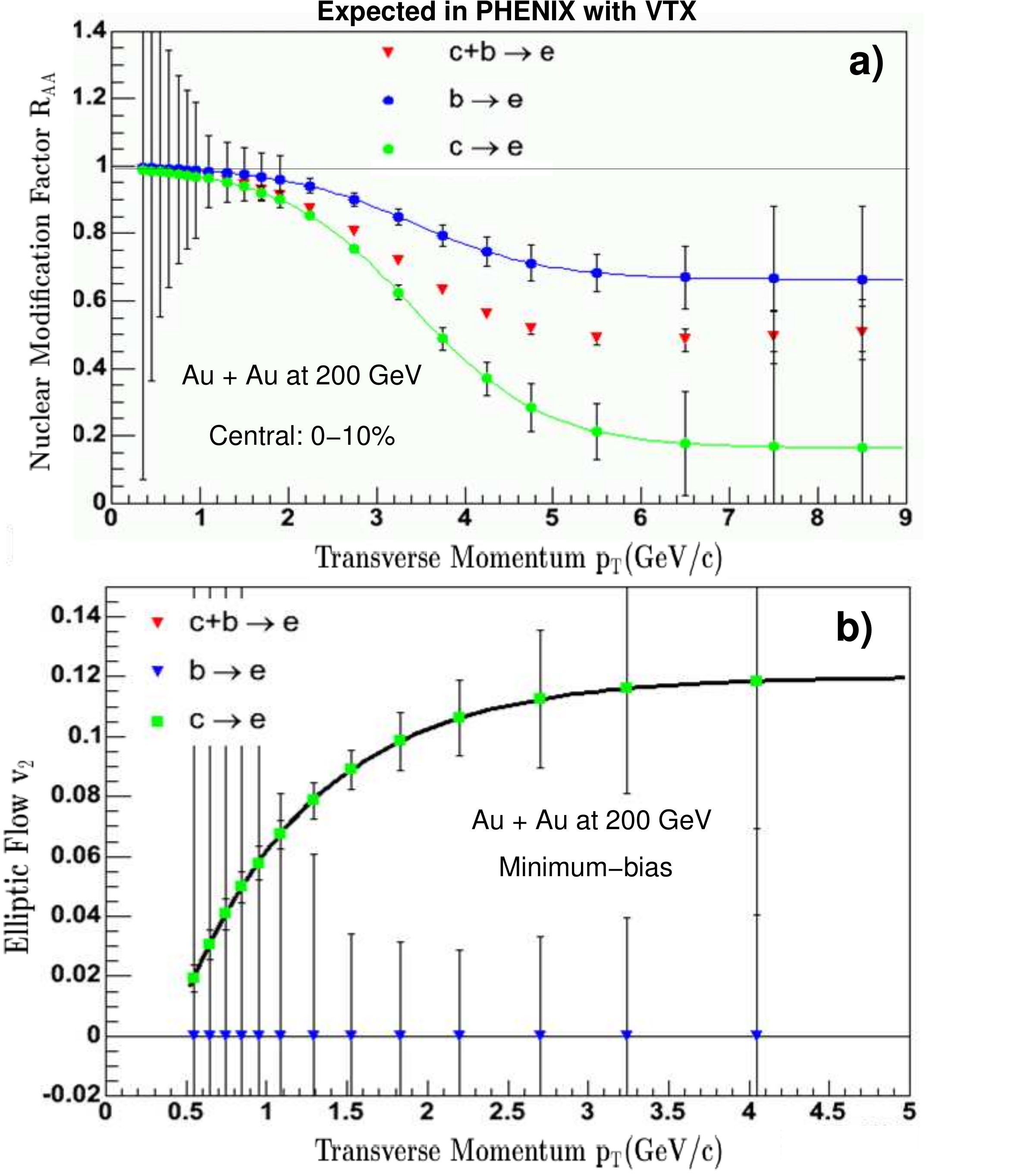}
\caption{\small Expected: a) nuclear modification factor, $R_{\rm
AA}$, and b) elliptic flow, ${\rm v_2}$, as a function of transverse
momentum of heavy-flavor electrons, $(e^{+} + e^{-}$)/2, from
heavy-flavor decays at midrapidity region in Au+Au collisions at \snn\
using PHENIX upgraded with silicon vertex tracker
detector (VTX)~\cite{Proposal}.
\label{fig:fig3}}
    \end{minipage}
\end{tabular}
\end{figure}
\vspace*{-0.4cm}
\section{Physics Motivations and Goals}
\vspace*{-0.2cm} Particles carrying heavy flavor, i.e. charm or beauty
quarks, are a powerful tool for study the properties of the hot and
dense medium created in high-energy nuclear collisions like at
RHIC. At the relatively low transverse momentum region, the collective
motion of the heavy flavor will be a sensitive signal for the
thermalization of light flavors at RHIC-II (RHIC-II is the upgrade of
the RHIC). They allow the probing of the spin structure of the proton
in a new and precise way.  \par The current PHENIX experiment at
Relativistic Heavy Ion Collider (RHIC) at Brookhaven National
Laboratory (BNL) is inadequately equipped to fully exploit the
opportunities of heavy-flavor production provides. Many of the
necessary measurements are either not possible or can be performed
only with very limited accuracy. Precise vertex tracking is imperative
for robust measurement of heavy-flavor production. The silicon vertex
tracker (covering $|\eta| <$ 1.2) which is under construction adds
tracking capabilities to the two central arms (each central arm covers
pseudorapidity region $|\eta| <$ 0.35 and is composed of the
drift chambers, pixel-pad chambers, and time expansion chambers) of
the PHENIX experiment.  With this detector, the charged particles detected
in the central arms can be identified as decay products from charm or
beauty carrying particles by measuring the distance of closest
approach (DCA) to the primary vertex in such events.

For heavy ion collisions or spin structure of the nucleon programs, we
plan to observe charm and beauty production through its semi-leptonic
decay to $e^{\pm}$. We will need a good vertex resolution to identify
the DCA to the primary vertex in such events. The main background
expected for this physics include Dalitz decays and photon conversions
have been studied using a GEANT detector simulation. We estimated that
the VTX detector could achieve $\sim$~40 ${\mu m}$ DCA resolution as
shown in figure~\ref{fig:fig1}.a. Using a DCA cut value of $\sim$ 200
$\mu m$ for tracks with ${\rm p_{_{T}} >}$~1~GeV/c, we should be able
to achieve a significant background reduction. As a result of the DCA
cut, the purity of the event sample increases from
$\sim$~50\%~to~$\sim$~90\% as presented by the simulation in
figure~\ref{fig:fig1}.b.

Some of physics topics addressed by the PHENIX experiment upgraded by
VTX detector in heavy ion collisions and spin structure of the nucleon
are~\cite{PHENIX1, PHENIX2}:
\subsection{Heavy Ion Collisions:}
The aim of ultra-relativistic heavy ion collisions is to produce and
reveal a new phase of matter. When ions collide at very high velocity
fascinating things happen. The RHIC is the first machine in the world
capable of colliding heavy ions at nearly speeds of light (what
Einstein called relativistic speeds). Recently, RHIC has discovered a
new state of matter strongly interacting and more like a ``perfect''
liquid explained by the equations of hydrodynamics~\cite{liquid}. The
evidence includes (1) bulk collective elliptic flow, (2) jet quenching
and mono-jet production, observed in Au+Au central collisions at 200
GeV per nucleon, and (3) a critical control experiment using d+Au
collisions at 200 GeV per nucleon. The research focus now shifts from
initial discovery phase towards detailed investigation of the
properties of the dense nuclear medium created in heavy ion
collisions. The PHENIX experiment upgraded by VTX
detectors~\cite{Proposal} will allow to establish very precise
measurements such as:
\begin{figure}[t]
\vspace*{-0.3cm}
\centering
\includegraphics[height=.27\textheight]{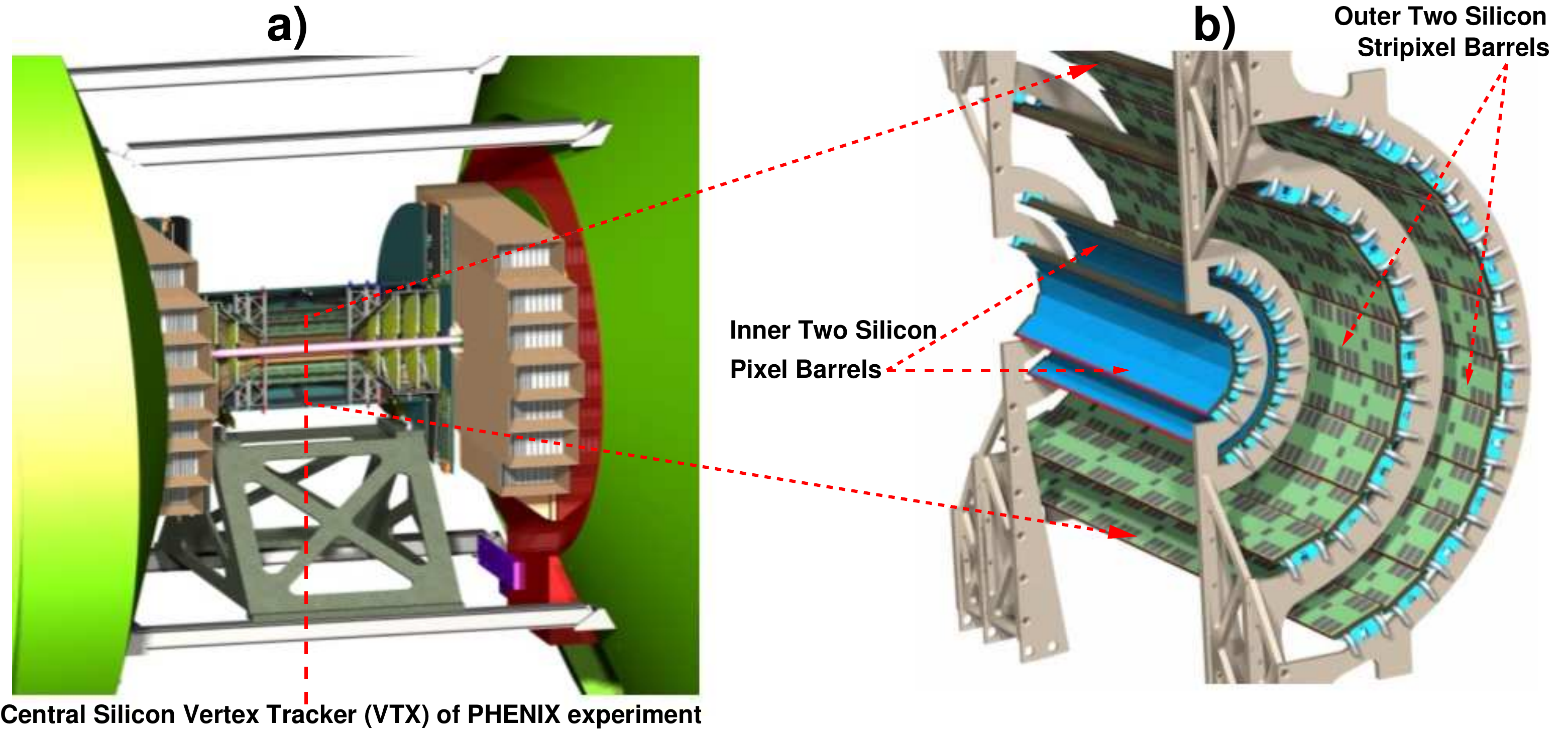}
\vspace*{-0.1cm}
\caption{\small Panel a) presents a 3-D view of central Silicon Vertex
Tracker (VTX) in PHENIX experiment\cite{Proposal}. Panel b) presents zoom on VTX
detector illustrating the two inner barrels of silicon pixel detector
and the two outer barrels of silicon stripixel detector.
\label{fig:fig4}}
\end{figure}
\begin{itemize}
\item mass dependence of the energy-loss of partons in the medium
expressed in figure~\ref{fig:fig2}.a by nuclear modification factor
$R_{\rm AA}$ relative to \pp collisions, which has been observed for
light partons, ($e^{+} + e^{-}$)/2. Figure~\ref{fig:fig3}.a shows the
expected results for charged particles ($e^{+} + e^{-}$)/2 detected in
the central arms identified as decay products from charm or
beauty in PHENIX experiment by adding VTX detector.
\item mass dependence of elliptic flow of quarks. 
Figure~\ref{fig:fig3}.b presents the elliptic flow amplitude of
electrons, ($e^{+} + e^{-}$)/2, issued from heavy-flavor decays at
midrapidity region in Au+Au collisions at \snn.
Figure~\ref{fig:fig2}.b presents the expected elliptic flow strength
v$_{2}$, of single electrons from heavy flavor decay measured in PHENIX
experiment by adding VTX detector. A strong elliptic flow is evidence
of strong coupling Quark-Gluon Plasma (sQGP).
\item a firm baseline to quantify the suppression or possible
enhancement of ${J/\psi}$
\end{itemize}
\par The present PHENIX can not distinguish single electrons from
charm decay and from beauty decay in Au+Au collisions. Thus we can not
determine the suppression factor, RAA (or elliptic flow, v2) of charm
accurately, nor we can determine if beauty also suffer significant
energy loss or not. The VTX detector will allow us to measure the
single electrons from charm and beauty separately. Since beauty has a
larger c$\tau$ (B$^0$: 462 $\mu$m, B$^+$: 502 $\mu$m) than charm (D$^0$: 123
$\mu$m, D$^+$: 317 $\mu$m), we can accurately split the beauty component
of single electron from the charm component using a precise displaced
vertex measurement from the VTX detector as shown in
Fig.~\ref{fig:fig3}. The VTX detector will also allow us to
measure the high-p${\rm{_T}}$ spectra of charm directly via the hadronic decay
channels, e.g. D $\rightarrow$ K $+$ $\pi$. From these measurements,
we will be able to determine the energy loss of charm and beauty in
the medium. This will be a decisive measurement to understand the
energy loss mechanism in the dense matter at RHIC.
\subsection{Spin Structure of the Nucleon:}
Spin plays a central role in our theory of the strong interaction,
 {Quantum Chromodynamics or QCD}. The PHENIX experiment with
 polarized protons colliding at RHIC probe the proton spin in new
 profound ways that are complementary to deep inelastic scattering
 experiments~\cite{Deep}. The proposed VTX detector will be crucial in the
 determination of gluon distribution in two significant ways:
\begin{itemize}
\item different measurements will cover the same kinematic regions:
this would enable the much-needed cross-checks within PHENIX for
accessing the polarized gluon distribution. The VTX detector extends
the reach in kinematic regions in ${x}$ (charm measurement and jet
reconstruction over large acceptance of VTX detector significantly
extend the $x$-range, where $x$ is the Bjorken variable which
describes the fraction of the nucleon momentum carried by a particular
parton) for many of the measurements and hence adds a significant
amount of overlap in ${x}$-range coverage. Here "coverage" implies we
measure the ratio $\Delta$G/G (gluon spin polarization) with ~20\%
relative uncertainty of its expected value at that x. For more details
see Ref.~\cite{Proposal}.
\item by being able to observe displaced vertices at 
low-${\rm p_{_{T}}}$ for semi-leptonic decays of charm and beauty, the VTX
detector will enable a large $x$-range over which we will make gluon
polarization measurements. It is estimated that the $x$ reach of the
VTX detector will be 0.01~$< x <$~0.3.
\end{itemize}
\begin{table}[t]
\vspace*{-0.3cm}
\caption {Summary of physical specifications of silicon vertex trackers (VTX)}
\label{tab:tab1}
\vspace*{-0.3cm}
\begin{center}
\begin{tabular}{|c|p{2cm}|p{1.5cm}|p{1.5cm}|p{2.5cm}|p{2.5cm}|}\hline
 & &\multicolumn{2}{c|}{Pixel Detector} &\multicolumn{2}{|c|}{Stripixel Detector}\\ 
\cline{2-6}
VTX & Layer & R1& R2 & R3 & R4 \\ 
\hline 
Geometrical  &R~(cm)         &  2.5 & 5    & 10    &   14  \\\cline{2-6}
dimensions   &$\Delta$z~(cm) & 21.5 & 21.8 &  31.8 &   38.2 \\\cline{2-6}
                       &Area~(cm$^{2}$)& 280  & 560  & 1960  &  3400  \\\cline{2-6}
\hline 
Channel         &Sensor size &\multicolumn{2}{c|}{1.28 $\times$ 1.36 }&\multicolumn{2}{c|}{3.43 $\times$ 6.36 } \\
count                     &R $\times$ z(cm$^2$)  &\multicolumn{2}{c|}{(256 $\times$ 32 pixels)}&\multicolumn{2}{c|}{(384 $\times$ 2 strips) } \\\cline{4-6}
\cline{2-4}

                     &Channel &\multicolumn{2}{c|}{50 $\times$ 425 $\mu$m$^{2}$}&\multicolumn{2}{c|}{80 $\mu$m $\times$ 3 cm} \\
                     &size    &\multicolumn{2}{c|}{ }&\multicolumn{2}{c|}{(effective 80 $\times$ 1000 $\mu$m$^{2}$)} \\\cline{2-6}
               & Sensor/ladder           & \multicolumn{2}{c|}{4 $\times$ 4 } &5 & 5  \\\cline{2-6}
               & Ladders                 &10            &20  &18   &26   \\\cline{2-6}
               & Sensors                 &160           &320 &90   &156   \\\cline{2-6}
               & Readout chips           &160           &320 &1080 &1872   \\\cline{2-6}
               & Readout channels        &1,310,720     &2,621,440 &138,240 &239,616 \\\cline{4-6}
\hline 
Radiation     & Total &\multicolumn{2}{c|}{1.44\%}&\multicolumn{2}{c|}{2.1\%} \\
length (X/X$_{0}$)         &       &\multicolumn{2}{c|}{      }& \multicolumn{2}{c|}{ }\\ \cline{4-6} 
\hline
Occupancy & Au+Au    & 0.53\%  & 0.16\%  & 4.5\% (x$-$strip) &  2.5\% (x$-$strip) \\
          & at 200 GeV      &         &         & 4.7\% (u$-$strip) &  2.7\% (u$-$strip) \\
\hline
\end{tabular}
\end{center}
\end{table}
\section{Overview of Silicon Vertex Tracker Detector} 
The central silicon vertex tracker (VTX) detector illustrated in
figure~\ref{fig:fig4} consists of four layers of barrel detectors,
covers almost 2$\pi$ azimuthal angle and pseudorapidity range ${\rm
|\eta|<}$ 1.2. The inner two layers are built from silicon pixel sensors
forming silicon pixel detector and the outer two layers are built from
silicon stripixel sensors forming silicon stripixel detector. 
The summary of physical specifications of VTX are summarized in
Table~\ref{tab:tab1}. 
The specifications have been chosen so that the design has a low
material budget to avoid tracks being randomly
bent by multiple scattering, and to minimize photon conversion (
$\gamma \rightarrow e^{+}e^{-}$) which generates a background for electron identification in outer
detectors. The VTX detector can be used in proton-proton collisions and the
high multiplicity environment of heavy ion collisions.
\begin{figure}[t]
\vspace*{-0.3cm}
\centering
\includegraphics[height=.35\textheight]{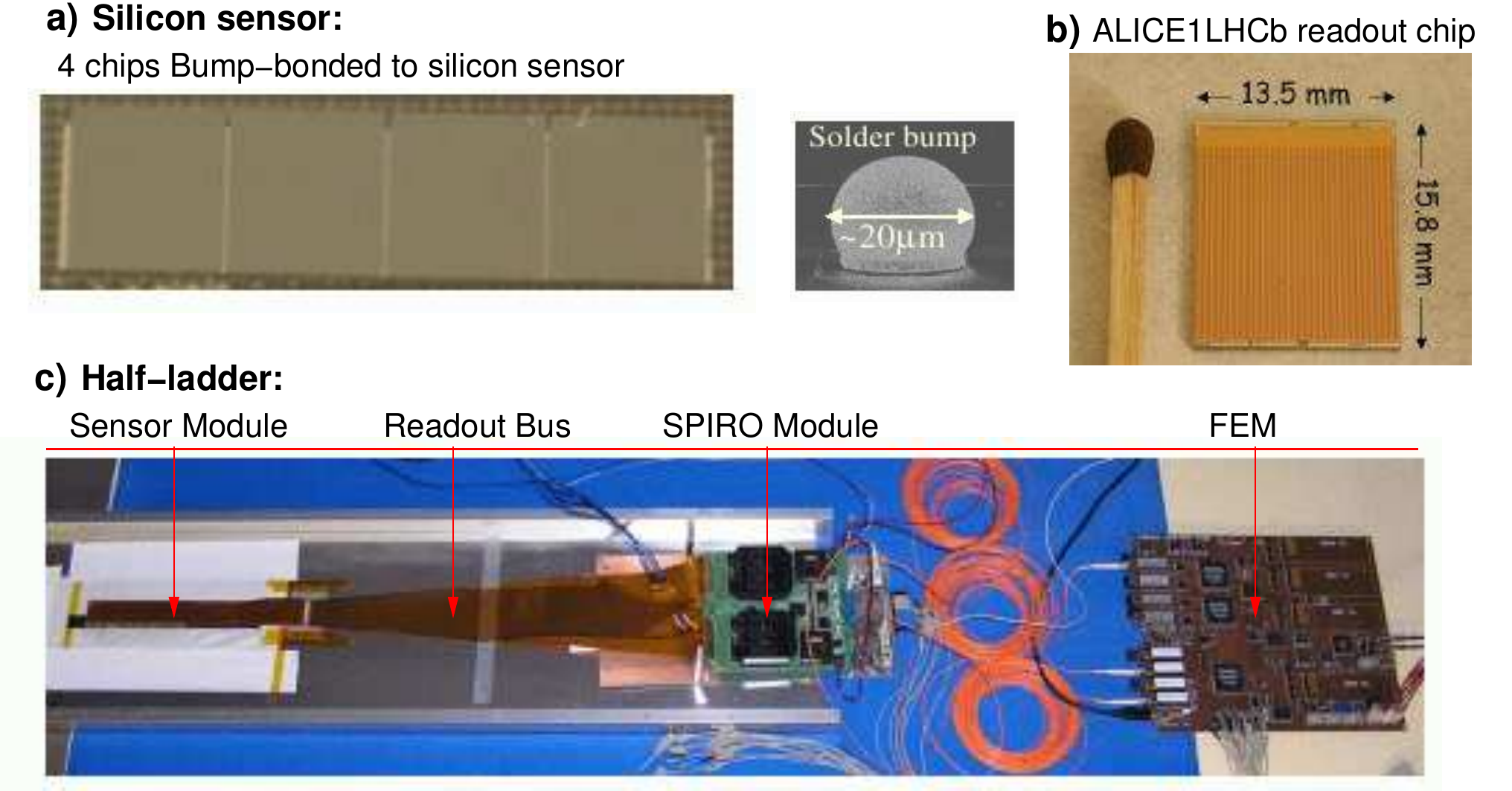}
\vspace*{-0.3cm}
\caption{\small Panels a), b) and c) show a picture of a silicon pixel sensor,
a picture ALICE1LHCb readout chip and a
picture of assembled half pixel ladder, respectively
\label{fig:fig5}}
\end{figure}
\begin{figure}[t]
\vspace*{-0.5cm}
\centering
  \begin{tabular}{cc}
    \begin{minipage}{2.8in}
\hspace*{-3.7cm}
\includegraphics[height=0.45\textheight]{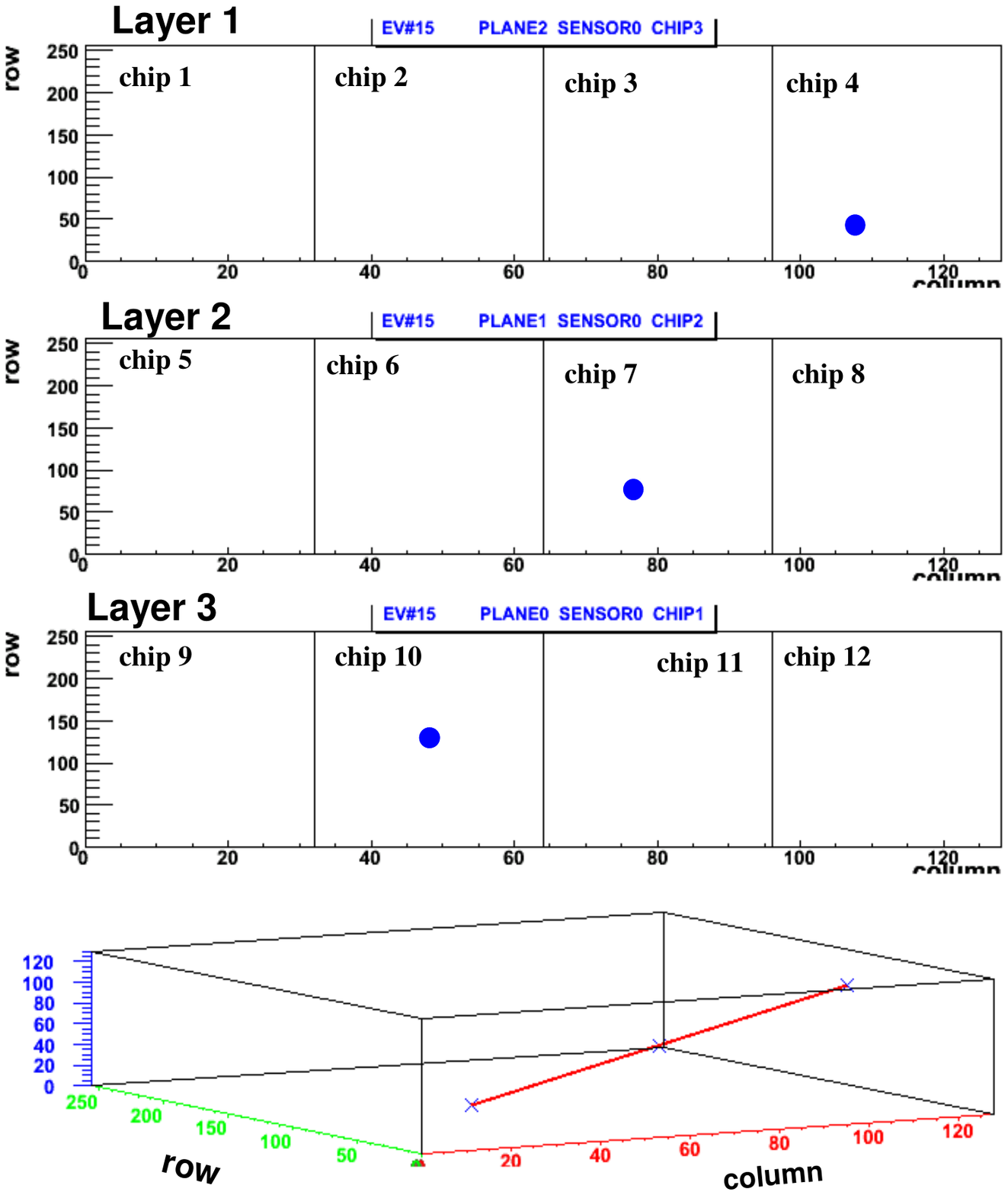}
\vspace*{-1.6cm}
\caption{\small Typical cosmic-ray event observed in the three layers
of silicon pixel detector (telescope).}
\label{fig:fig6}
\end{minipage}
&
\begin{minipage}{3in}
\hspace*{-1.0cm}
\vspace*{-1.3cm}
\includegraphics[height=0.4\textheight, angle=-90]{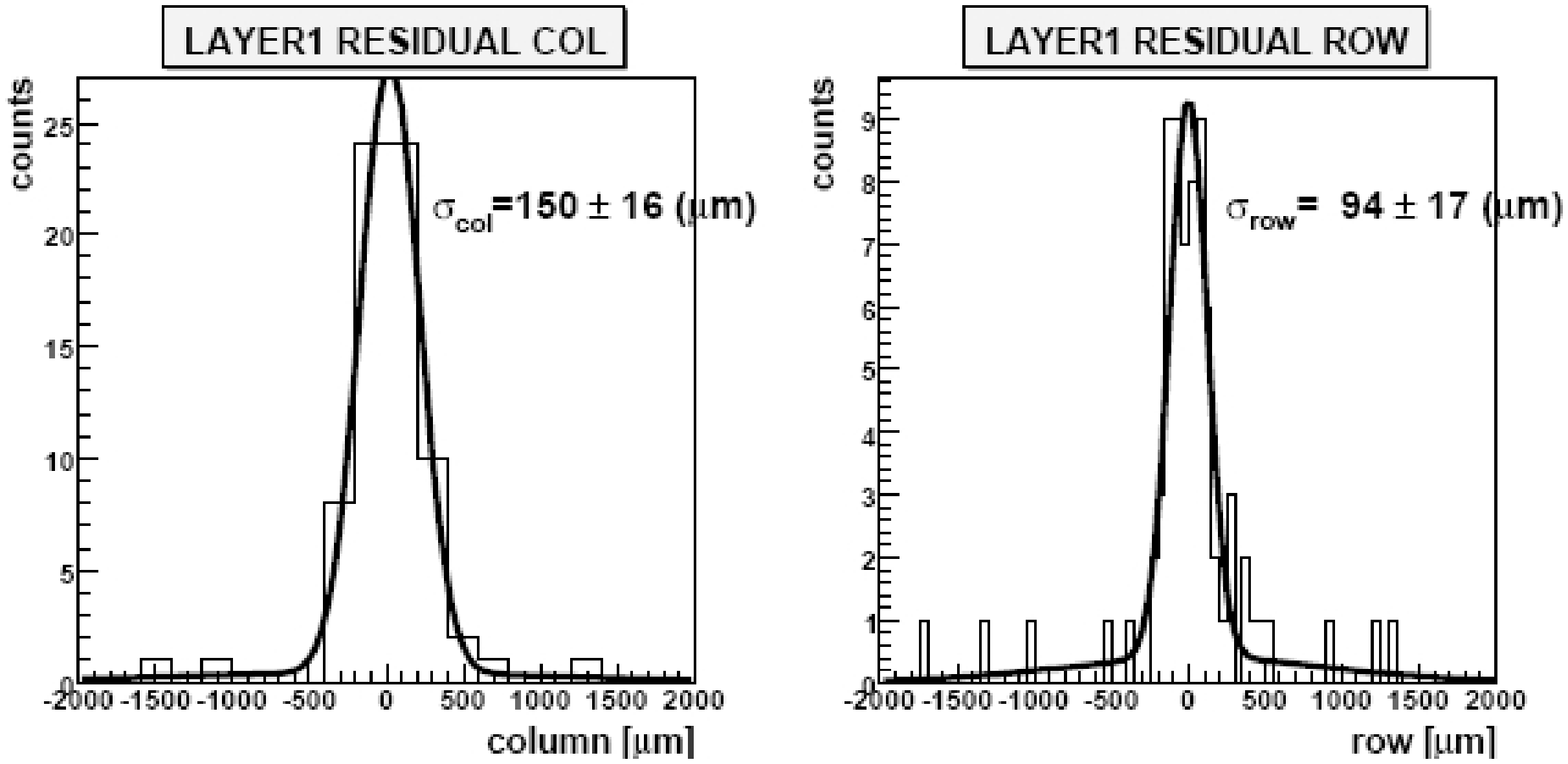}
\vspace*{-0.5cm}
\caption{\small Residual distributions for tracks
in the cosmic ray sample: (left panel) the $z$-side (column pixels)
and (right panel) the $\phi$-side (row pixel).\label{fig:fig7}}
  \end{minipage}
\end{tabular}
\end{figure}
\subsection{Silicon Pixel Detector} 
The two inner barrels of VTX detector consists of silicon pixel
detectors (SPD)~\cite{Proposal}. The technology of SPD is similar to
the ALICE1LHCb sensor-readout hybrid, which was developed at CERN for
the ALICE and LHCb experiment~\cite{ALICE1, ALICE2}. The physical
specifications of silicon pixel detector are summarized in
Table~\ref{tab:tab1}. The first and second silicon pixel barrels
consist of 10 and 20 ladders, respectively. One ladder is electrically
divided into two independent half-ladders. Each ladder consists of
four sensor modules mounted on a mechanical support (stave). Each
sensor module consists of a silicon pixel sensor bump-bonded with four
pixel readout chips. The readout chips (ALICE1LHCb) were developed by
the CERN EP-MIC group and fabricated by using the IBM 0.25~$\mu$m CMOS
technology~\cite{ALICE3}. The library and the design techniques of the
readout-chips guarantee the radiation hardness of the chip up to 30
kRads (for more details see Ref.~\cite{ALICE3}). The size of the chip
is 15.6 mm $\times$ 13.7 mm with 150~$\mu$m thickness. One chip has 32
$\times$ 256 = 8,192 channels (pixels), each channel is pre-amplified
and discriminated with a configurable threshold level. The
discriminated binary signal is delayed with programmable duration, and
then, stored to be read out by a downstream data acquisition
system. The chip has 44 internal 8-bit Digital-to-Analog Converters
for controlling the threshold of discriminators and various
timings. All configurations are set via the JTAG serial interface
(IEEE std. 1149.1-1990). The chip is operated using a 10~MHz
clock. Therefore, the maximum readout speed is 25.6 $\mu$sec with a
32-bit line. The sensor chip thickness is 200~$\mu$m and its
fabricated by CANBERRA. It is partitioned into four active areas. Each
area has 32 columns (z) $\times$ 256 rows ($\phi$) with a pixel size
of 425 $\mu$m (z) $\times$ 50 $\mu$m ($\phi$). Pictures of silicon
pixel module, pixel readout chip and the structure of half-ladder are
presented in figure~\ref{fig:fig5}. a), b) and c), respectively.  \par
The sensor modules of one half-ladder are wire-bonded to a readout bus
made of Copper-Aluminum-Polyimide Flexible Printed Circuit board
(FPC), which is supported by a carbon composite stave. This basic
detector (half ladder) unit is read-out by a SPIRO module (SPIRO:
Silicon Pixel Intermediate Read-Out). SPIRO modules are placed outside
of the detector acceptance. They process the incoming control signals
and transmit the outgoing data of a pixel half ladder. A SPIRO module
carries analog Pilot chips for the power and reference voltage
supplies of the pixel readout chips, digital Pilot chips for their
controls and readout, and an optical link chips and transmitters for
the data transfer. The SPIRO modules are then connected to pixel Front
End Modules (FEMs) outside of the PHENIX IR. The FEMs work as
interface to the PHENIX DAQ system. Since the SPIRO module transmits
the data via optical fiber, the FEM can be located in electronic racks
away from the vertex region. Each FEM may receive data from several
pixel half ladders and thus reduce the number of Data Collection
Modules (DCM) needed to interface to the PHENIX DAQ. In order to allow
simple manipulations of the data, the FEM will pipe the data through
an FPGA. This FPGA will add data headers and trailers to for standard
PHENIX data packages. The design of the FEM is very similar to FEM's
that are currently employed in the PHENIX readout system. Additional
information can be found in Ref.~\cite{Eric}.
\subsubsection{Performance of the silicon pixel detectors} 
To ensure long-term operation of silicon pixel detector, it is
mandatory to study their performance. Recently, the functionality of
the prototype silicon pixel detectors with pixel readout board (SPIRO)
using PHENIX readout system was performed using cosmic-ray test~\cite{Kurosawa}.  The
setup consists of a telescope made from three pixel layers 
(three half-ladders) connected to the PHENIX
DAQ system. The trigger is made of a coincidence of self trigger
signal (FastOR) on the layer-1, layer-2 and
layer-3. Figure~\ref{fig:fig6} shows typical cosmic-ray event observed
by the three silicon pixel layers. Figure~\ref{fig:fig7} shows the
residual distributions on the $\phi$-side and $z$-side for tracks in
the cosmic-ray sample.  \par The prototype silicon pixel detector was
confirmed to be well functioning by finding tracks reconstructed by a
telescope. The performance of the bus is measured and is compared
with the simulation. The comparison with simulation indicates that the
functionality of the pixel ladders satisfy the PHENIX requirements
for the upgrade~\cite{Proposal}.
\subsection{Silicon Stripixel Detector} 
The outer two barrels of the VTX detector for PHENIX experiment
upgrade consists of silicon stripixel detector with a new "spiral"
design, single-sided sensor with 2-dimensional (X-U) readout~\cite{Proposal}.
\subsubsection{ Novel Stripixel Sensor Design and Specifications}
\begin{figure}[t]
\vspace*{-0.3cm}
\centering
  \begin{tabular}{cc}
    \begin{minipage}{2.8in}
\hspace*{-1.5cm}
\includegraphics[height=0.55\textheight]{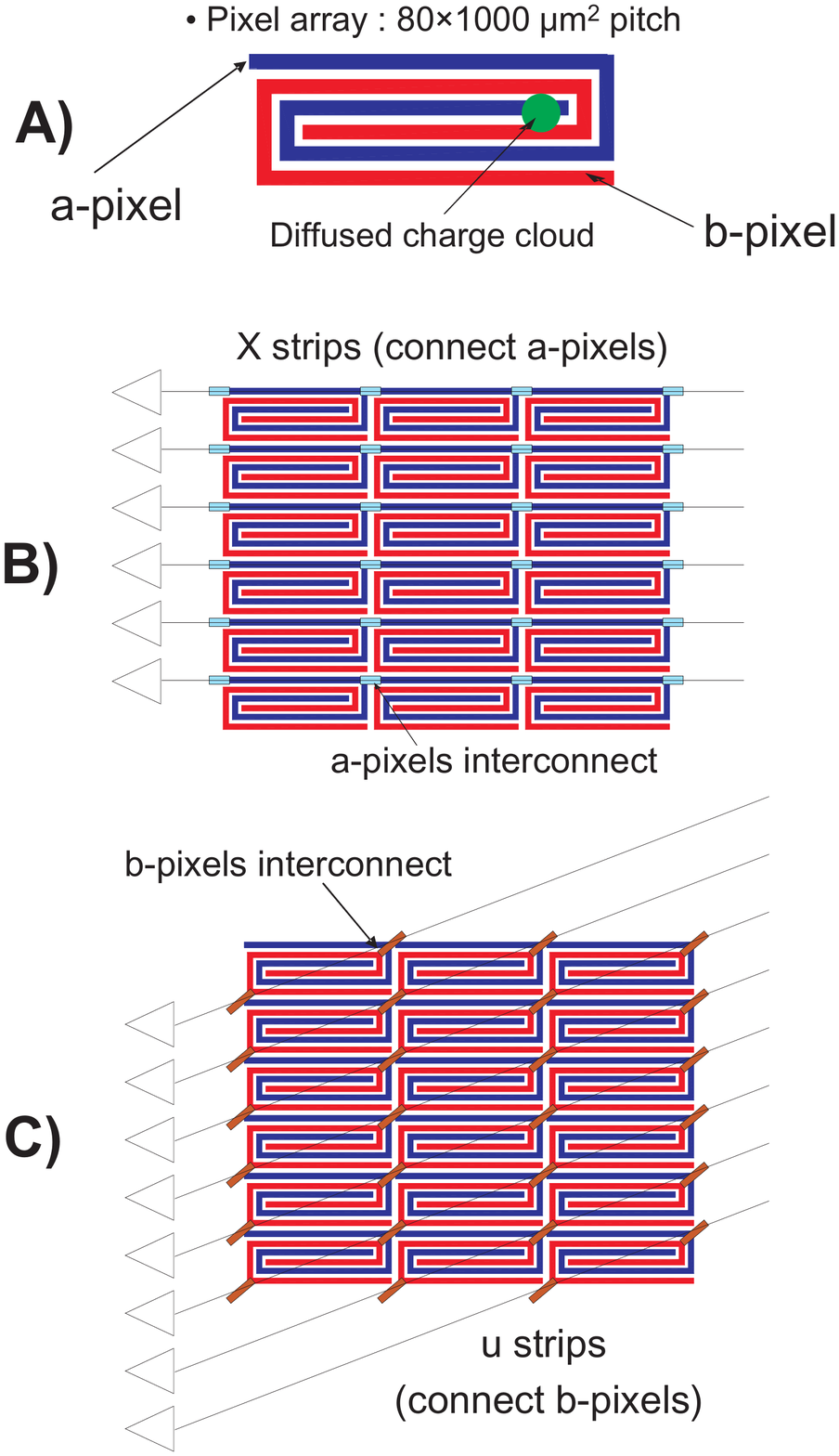}
\caption{\small Panel (a) stripixel sensor concept with two independent interleaved
spiral shaped a-pixel and b-pixel. Panel (b) a-pixels are connected in such
way to form a X-strips. Panel (c) b-pixels are connected to form a U-strips~\cite{Rachid}}
\label{fig:fig8}
    \end{minipage}
&
\begin{minipage}{2.8in}
\hspace*{0.2cm}
\vspace*{-0.3cm}
\includegraphics[width=0.9\textwidth]{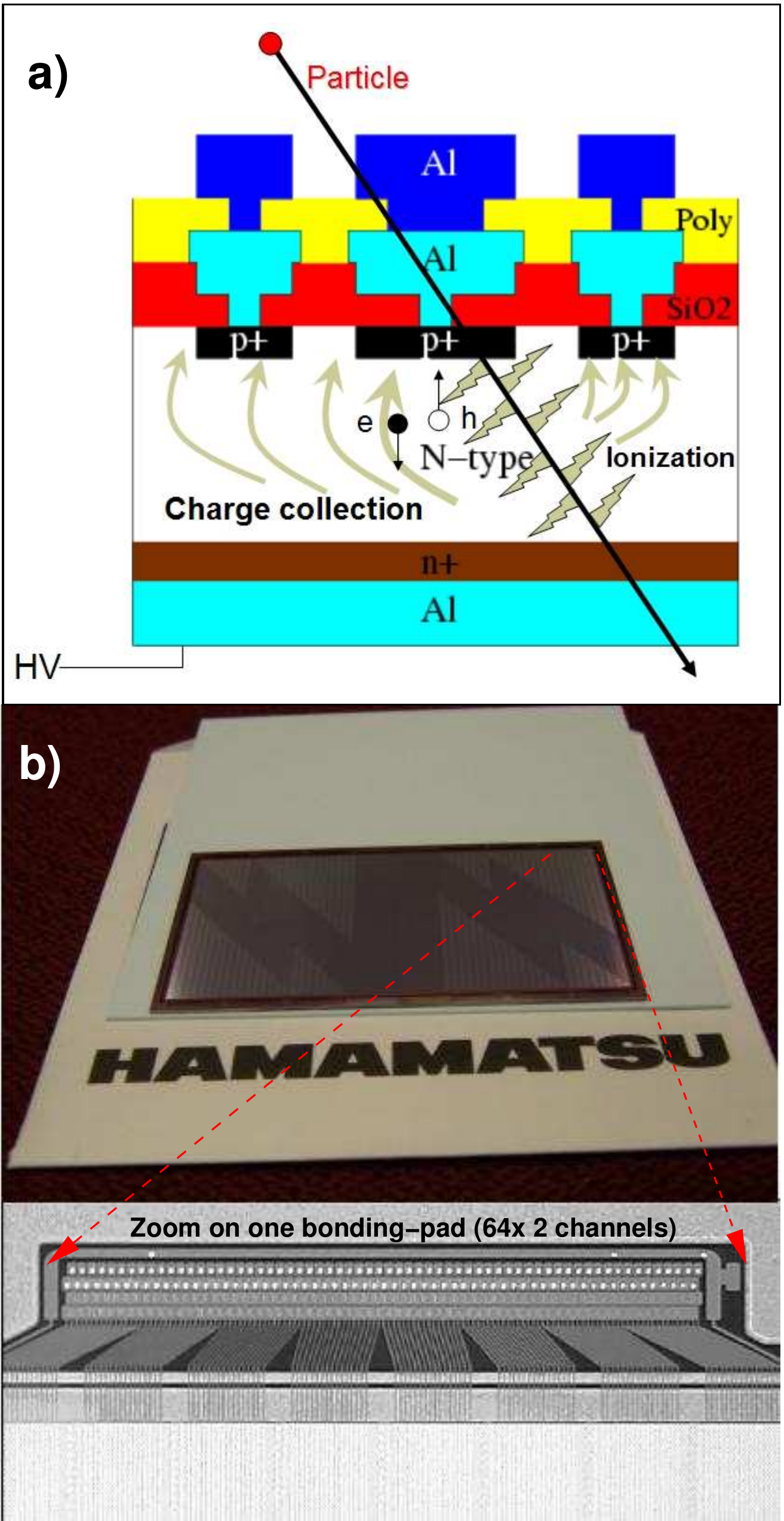}
\hspace*{-0.4cm}
\caption{\small Panel a) Cross section view of double metal layout of
silicon stripixel sensor via contacts on b-pixels of
U-strip~\cite{Rachid}. Panel b) Photo of one silicon stripixel sensor from
full-production fabricated by HPK.}
\label{fig:fig9}
  \end{minipage}
\end{tabular}
\end{figure}

A novel stripixel silicon sensor has been developed at
BNL~\cite{Zheng}. The silicon sensor is a single-sided, DC-coupled,
two-dimensional (2D) sensitive detector. This design is simpler for
sensor fabrication and signal processing than the conventional
double-sided strip sensor. Each pixel from the stripixel sensor is
made from two interleaved implants (a-pixel and b-pixel) in such a way
that the charge deposited by ionizing particles can be seen by both
implants as presented in figure~\ref{fig:fig8}.A. The a-pixels are
connected to form a X-strip as is presented in
figure~\ref{fig:fig8}.B. The b-pixels are connected in order to form a
U-strip as is presented in figure~\ref{fig:fig8}.C. The stereo angle
between a X-strip and a U-strip is 4.6$^{o}$.  A schematic cross
section of the silicon stripixel sensor is presented in
figure~\ref{fig:fig9}.a. The basic functionality of the sensor is
simple; signal charges (electron$-$hole pairs) generated for example
by particles produced from collisions are separated by the electric
field, the electrons moving to the n$^{+}$ side, holes to the p$^{+}$
side, thus producing an electric signal which can be amplified and
detected. In figure~\ref{fig:fig9}.a, the first Al layer is the metal
contacts for all pixels. All X-strips are routed out by the first
metal Al layer. All U-strips are routed out by the second metal Al
layer.  \par The size of the silicon stripixel sensor is about 3.43
$\times$ 6.36 cm$^{2}$ and is shown in figure~\ref{fig:fig9}.b. In
each long side of the sensor there are six sections of bonding pads,
with 128 bonding pads each. This implies that each sensor has 2
$\times$ 3 $\times$ 128 = 768 of X-strips of 80 $\mu$m width and 3.1
cm length in beam direction and the same number of U-strips at an
angle of 4.6$^o$ to the beam direction. Due to the stereoscopic
readout the effective pixel size is 80 $\times$ 1000 $\mu$m. Five (for
layer 3) or six (for layer 4) sensors are mounted in a ladder. The
full length of a ladder in the beam direction is 31.8~cm (for layer 3)
or 38.2 cm (for layer 4). A total of 44 ladders are required to cover
the azimuth acceptance almost 2$\pi$. The geometric characteristics of
silicon stripixel layers are presented in Table~\ref{tab:tab1}. The
novel stripixel silicon sensor technology developed, including the
mask design and processing technology, has been transferred from BNL
to sensor fabrication company Hamamatsu Photonics (HPK) located in
Japan, for mass production. A picture of one sensor from full
production is presented in figure~\ref{fig:fig9}.b.
\begin{figure}[t]
\vspace*{-0.3cm}
\centering
  \begin{tabular}{cc}
    \begin{minipage}{2.9in}
\vspace*{-0.3cm}
\hspace*{-0.0cm}
\includegraphics[height=0.23\textheight]{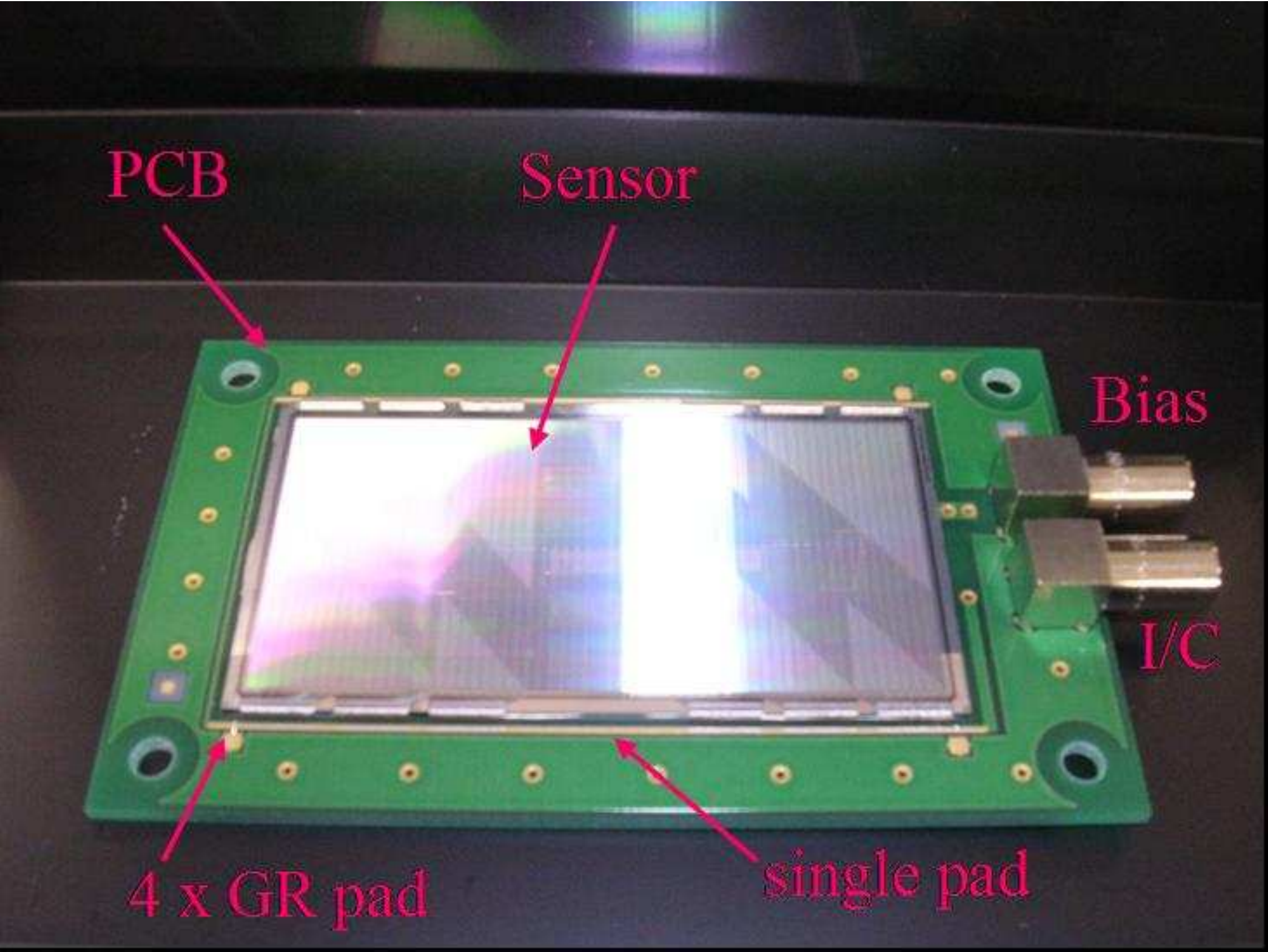}
\caption{\small Picture of sensor 625 $\mu$m thickness mounted
on PCB to apply bias voltage. All strips were wire bonded to a single
readout on the PCB. The guard ring of the sensor was wire bonded to
the GND connection of the PCB.
\label{fig:fig10}}
  \end{minipage}
&
    \begin{minipage}{2.7in}
\vspace*{-0.5cm}
\hspace*{-0.3cm}
\includegraphics[width=1\textwidth]{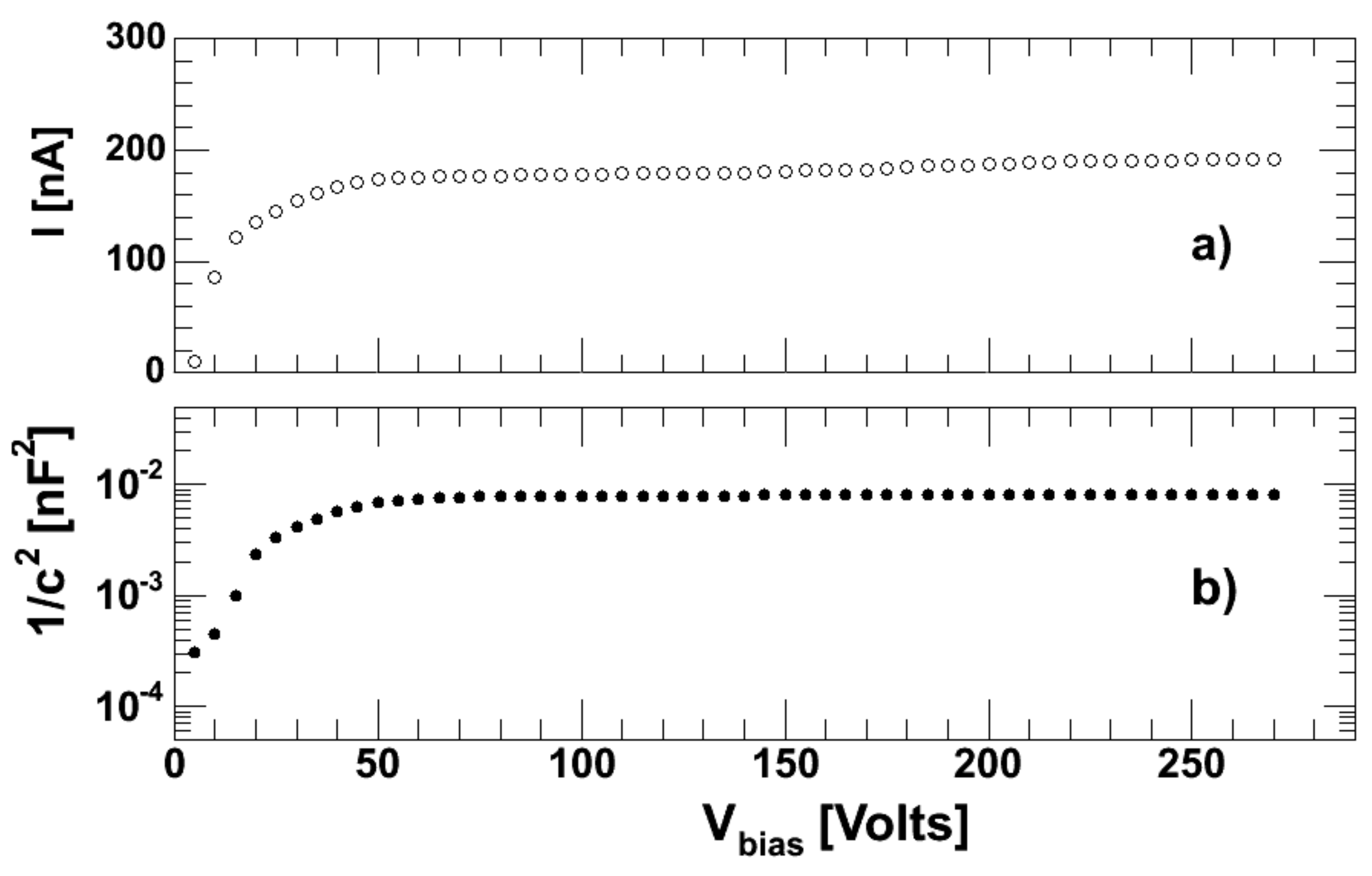}
\caption{\small Panel a) total leakage current obtained from stripixel sensor
presented as a function of bias voltage. Panel b) capacitance distribution
obtained in the similar conditions as panel~a). For 
additional information on measurements can be found in Ref.~\cite{Rachid}.\label{fig:fig11}}
    \end{minipage}
\end{tabular}
\end{figure}
\par To measure the total leakage current (I${\rm _{tot}}$) of the
sensor and eventually to extract the leakage current per strip (I${\rm
_{strip}}$ = I${\rm _{tot}}$/N${\rm _{tot}}$ where the total number of
strips N${\rm _{tot}}$ = 12 sections $\times$ 128 strips = 1536
strips), two sensors that passed the quality assurance tests were
mounted on two independent print circuit (PCBs) and wire bonded as
shown in figure~\ref{fig:fig10}. The current and capacitance as a
function of bias voltage were measured and normalized to 20 $^o$C and
they are presented in figure~\ref{fig:fig11}. We have studied the
stability of the leakage current as function of time for a long period
(22 days) and the temperature has been monitored for the same
period. We observe that the leakage current is stable as a
function of time and it has good correlation with the temperature of
the clean-room~\cite{Rachid}. In figure\ref{fig:fig11}, the measured
total leakage current per stripixel can be obtained as I${\rm
_{tot}}$~=~179~nA this implies I${\rm _{strip}}$ = 179/1536 strips =
0.12~nA which is very low and it allows the use of the SVX4 chips,
which have a hard limit of 15 nA/strip. This hard limit comes from
that the leakage current from the DC-coupled sensor will rapidly
saturate the SVX4 input preamp. The preamp dynamic range is 200 fC,
which will saturate in 500 $\mu$s with a 0.4 nA/strip.  However, we
can issue a preamp-reset signal once per RHIC abort ($\sim$ 13
$\mu$s). In fact, this reset frequency limitation puts a hard limit on
the maximum acceptable leakage current. If the leakage current stays
below this limit, we should not expect any problems due to leakage
current~\cite{Rachid}.
\subsubsection{Performance of Prototype Silicon Stripixel Module using ROC-V2} 
\begin{figure}[t]
\centering
\vspace*{-0.3cm}
\includegraphics[height=.45\textheight]{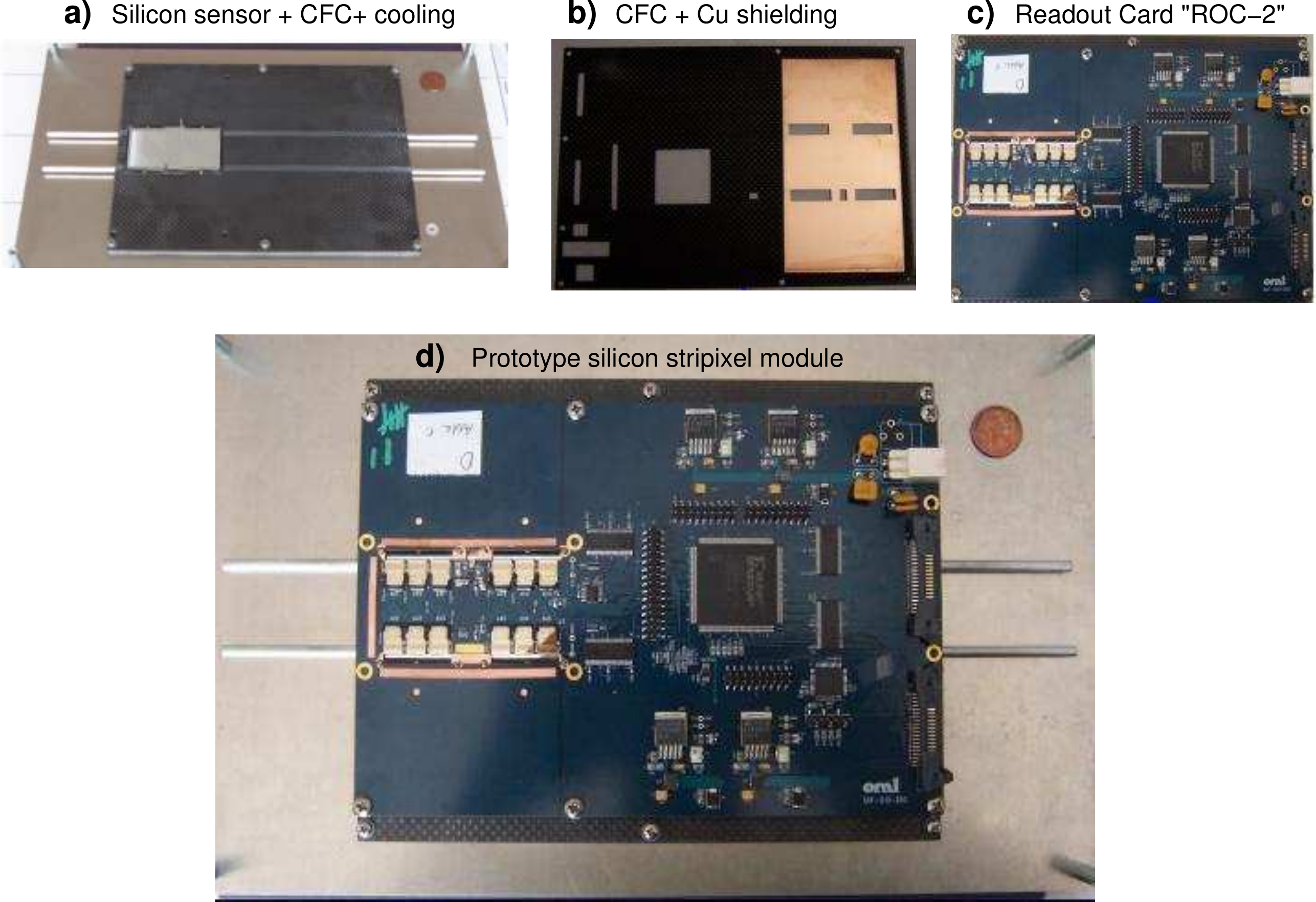}
\vspace*{-0.3cm}
\caption{\small Pictures of prototype silicon
 stripixel module components: a) one sensor
 thermal plate, b) one shielding layer and c) one Read-out Card version
 two (ROC-2).\label{fig:fig12}}
\end{figure}

A prototype silicon stripixel module has been fabricated to evaluate
performance of the stripixel sensor and Readout Card Version two
(ROC-V2). As it is shown in figure~\ref{fig:fig12}.a), b) and c), the
silicon stripixel module consists of one sensor thermal plate, one
shielding layer and one Readout Card Version two (ROC-V2),
respectively. The sensor thermal plate consists of one silicon sensor
mounted on thermal plate made from Carbon Fiber composite (CFC) that
contains two cooling tubes. The sensor thermal plate has also four
spacers used to elevate the shielding layer and ROC-V2 from touching
the sensor. The shielding layer consists of one CFC layer and one
Copper layer (18 $\mu$m thickness). The ROC-V2 is a thin printed
circuit board (600 $\mu$m) that hosts twelve SVX4s readout chips plus
one field-programmable gate array (FPGA) required to digitize the data
from one stripixel sensor and a few passive components.
\par The stripixel sensor is read out with the SVX4 chip developed by
FNAL/Berkeley collaboration.  The SVX4 chip is designed and fabricated
in the 0.25 $\mu$m CMOS process on 300 $\mu$m thick silicon and has
been successfully tested up to 10 Mrads. It contains 128 parallel
charge integration channels and 8-bit ADCs. The chip is designed to
run in a deadtime-less mode: the front-end (analog) part can run in
parallel with the back-end (digital) part. For each channel, there are
46 pipeline cells to store the data. These can hold at most 4 events
for further digitizing and readout. The operation voltage of the chip
is 2.5 V. The other features of SVX4 chip are: 1) max interaction rate
at 132 ns, 2) optimized capacitance load between 10 to 35 pF, 3)
built-in charge injection for preamp calibration, 4) channel mask to
exclude channels with excessive DC current input during operation, 5)
readout in byte-serial mode with optional zero suppression
(sparsification), 6) real-time event-by-event pedestal subtraction
(RTPS) which can suppress the common mode noise and compensate for any
residual pickup effects and 7) adjustable and loadable control
parameters. The noise of a SVX4 chip is around 400 electrons. The
dynamic range of the amplifier is 200 fC. The threshold and gain of
the amplifier can be set to effectively readout a minimum ionizing
particle (MIP) signal. Additional information on the SVX4 can be found
in Ref.~\cite{SVX4}. Prototyping efforts have allowed us to verify
compatibility of the SVX4 with the PHENIX DAQ (serial programming,
clock, fast control and data read-out). Command parsing and data
formatting functions are implemented in FPGA and have been largely
exercised with existing prototypes developed at ORNL (Oak Ridge
National Laboratory). The data formatting code is particularly simple,
and is almost entirely a subset of code developed for previous PHENIX
subsystems. The interface between ROC-2 and the data acquisition (DAQ)
is done by a Front-end Module (FEM) prototype. This communicates to a
PC-computer via USB interface.  \par For the evaluation of the
performance, the silicon stripixel module shown in
figure~\ref{fig:fig12}.c was placed inside a dark-box and connected to
the cooling system.  The cooling system was running at 0~$^o$C degree
temperature. The cooling at 0 $^o$C temperature is required when the
detector will be operational under high radiation exposure (total
fluence integrated over 10 years amounting up to 3.3 $\times$
10$^{12}$ n-eq/cm$^2$) at RHIC-II and we have to keep the leakage
current of the sensor below the saturation limit of SVX4, 15
nA/strip~\cite{Rachid, SVX4}. The temperature and humidity inside the
dark-box were monitored and were relatively stable as a function of
time.  \par
\begin{figure}[t]
\centering
\vspace*{-0.3cm}
\vspace*{-0.3cm}
\includegraphics[height=.22\textheight]{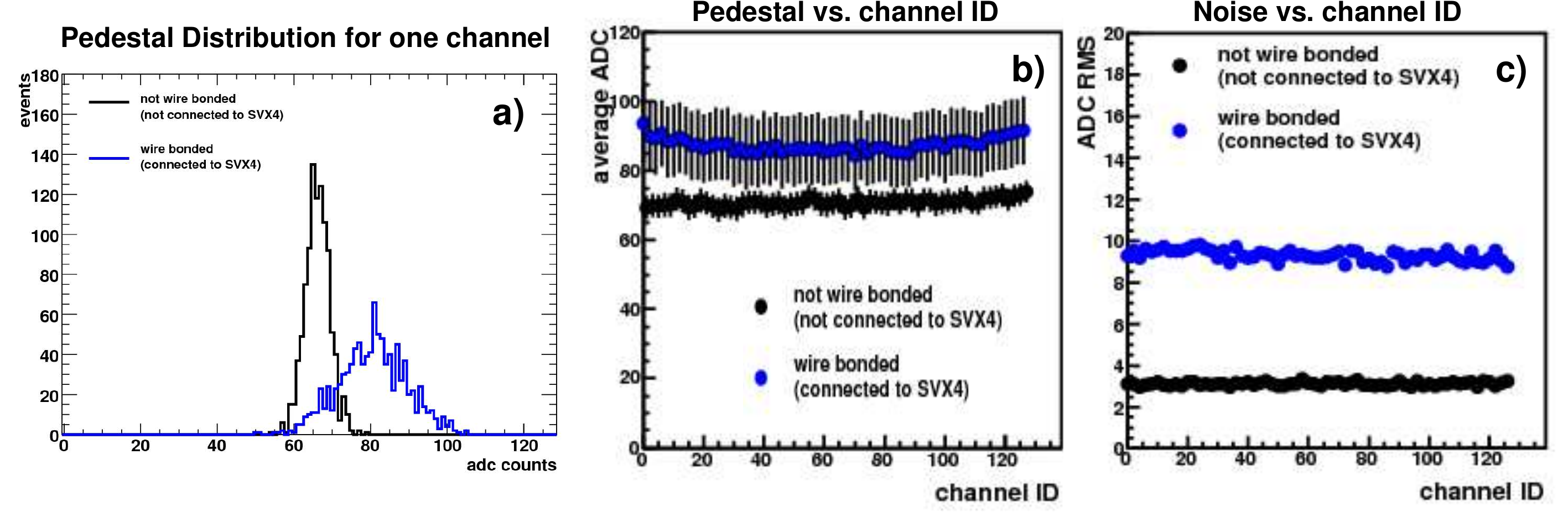}
\vspace*{-0.5cm}
\caption{\small Test results of silicon stripixel module using ROC-V2:
a) pedestal distribution for a given channel, b) pedestal versus
channel number for one section (one SVX4 chip) of silicon stripixel
module and c) noise versus channel number for one section (one SVX4
chip) of silicon strip module. \label{fig:fig13}}
\end{figure}

 The primary goal of the evaluation of silicon stripixel
module using ROC-V2 is to study the flatness of the pedestal and noise
distributions as a function of stripixel number as well the stability
of the distributions event-by-event. These informations are crucial
for the use of the zero-suppression of SVX4 when it is connected to
the silicon sensor and the configuration of the {ROC-V2}. The silicon
stripixel module performances for the signal-to-noise and effects of
irradiation dose on the silicon leakage current have been already done
successfully by the first prototype silicon modules and can be found
in Refs.~\cite{Proposal, Rachid, Tojo}. The pedestal distribution for
a given channel, pedestal and noise distributions versus channel
number of silicon stripixel module are presented in
figure~\ref{fig:fig13}.a , b and c, respectively. For the silicon
stripixel module which used spacers to elevate the ROC-V2 from
touching the silicon sensor, we did not succeed to do the full
wire-bonding of all SVX4s to the silicon sensor (we had technical
difficulties in wire-bonding the inner bonding pads of the silicon
sensor to the SVX4s). However, by removing the spacers we succeeded to
have one full wire-bonded silicon stripixel module (full wire-bonding
of the twelve SVX4 to the silicon sensor). From test results, we
observe that the noise per channel of wire-bonded channel of silicon
stripixel module using ROC-V2 is around 9 ADC units and the noise
distribution versus stripixel channel is flat. However, we observed
that pedestal distribution of wire-bonded channels versus channel
number are not completely flat.  The test results of the silicon
module using ROC-V2 suggest that there is charge sharing between the
analog front ends of the SVX4 readout chips. Improved grounding by the
inclusion of a solid, continuous, uniform analog-ground plane under
the SVX4s and sensor, along with improved connections between the
analog-ground and the sensor shield will reduce the amount of charge
sharing.
\section{Summary}
The physics capabilities in heavy ion collisions and spin structure of
the nucleon programs added to PHENIX by the new central silicon vertex
tracker (VTX) upgrade have been presented. The technology choices used
in the design, specifications, assembly procedures and data readout of
prototype silicon modules of VTX, pixel and stripixel modules, have
been elucidated. Cosmic-ray test results confirmed that prototype
silicon pixel telescope (telescope made from three pixel layers) are
performing well by finding tracks reconstructed. The performance of
the pixel ladders was measured and the comparison with simulations
indicates that the functionality of the pixel detector satisfy the
PHENIX experiment requirements for the upgrade. The test results of
the silicon module using ROC-V2 suggest that there is charge sharing
between the analog front ends of the SVX4 readout chips. Improved
grounding by the inclusion of a solid, continuous, uniform
analog-ground plane under the SVX4s and sensor, along with improved
connections between the analog-ground and the sensor shield will
reduce the amount of charge sharing. The present status of the silicon
stripixel module, we are in process of making new ROC-V3 which
includes all informations obtained from silicon stripixel module using
ROC-V2. We expect to have the first test results of the silicon
stripixel module using the ROC-V3 readout card in first half of 2008.
\vskip 0.5cm
\noindent{\bf \small Acknowledgments:\\} 
{This manuscript has been authored by employees of Brookhaven Science
Associates, under Contract No. DE-AC02-98CH10886 with the
U.S. Department of Energy.
}

\end{document}